\documentstyle[aps,aps12,epsf]{revtex}

\begin{document}

\title{\vspace*{-1cm}\hfill\mbox{\small FSU-SCRI-98-128}\\
\vspace*{1cm}A study of chiral symmetry in
quenched QCD using the Overlap-Dirac operator\\[+0.5cm]}

\author{
Robert G. Edwards, Urs M. Heller, Rajamani Narayanan}
\address{
SCRI, Florida State University, 
Tallahassee, FL 32306-4130, USA\\
e-mail: \{edwards,heller,rnaray\}@scri.fsu.edu}

\maketitle 

\vskip 1cm

\begin{abstract}

We compute fermionic observables relevant to the study of chiral
symmetry in quenched QCD using the Overlap-Dirac operator for a wide
range of the fermion mass.  We use analytical results to disentangle
the contribution from exact zero modes and simplify our numerical
computations.  Details concerning the numerical implementation of the
Overlap-Dirac operator are presented.

\end{abstract}

\section{Introduction}
\label{sec:intro}

The Overlap-Dirac operator~\cite{Herbert1} derived from the overlap
formalism for chiral fermions on the lattice~\cite{over} provides a
non-perturbative regularization of chiral fermions coupled vectorially
to a gauge field. The massless Overlap-Dirac operator is given by (the
lattice spacing is set to one)
\begin{equation}
D(0)={1\over 2} \left[1 + \gamma_5\hat H_a \right].
\label{eq:D0}
\end{equation}
$\hat H_a$ is a hermitian operator that depends on the background gauge field
and has eigenvalues equal to $\pm 1$. There are two sets of choices for
$\hat H_a$. The simplest example in the first set is $\epsilon(H_w)$ 
where $\gamma_5 H_w(-m)$ ($m_c < m < 2$) 
is the usual Wilson-Dirac operator on the 
lattice with a negative fermion mass~\cite{Herbert1} and $-m_c$ is the
critical mass, which goes to zero in the continuum limit. One could replace
$H_w$ by any improvements of the Wilson-Dirac operator with the mass in the
supercritical region but smaller than where the doublers become light.
The simplest example in the other set is obtained from the domain wall 
formalism~\cite{kaplan} and $\hat H_a=\epsilon(\log(T_w))$ where $T_w(m)$ is
the transfer matrix associated with the propagation of the four
dimensional Wilson fermion in the extra
direction~\cite{Herbert3}. Here $m_c < m < 2$ is the domain wall mass. 
Again, one can replace the five dimensional Wilson-Dirac fermion 
by any improvements.
The massive Overlap-Dirac operator
is given by
\begin{equation}
D(\mu)={1\over 2} \left[1+\mu + (1-\mu) \gamma_5\hat H_a \right]
\label{eq:Dmu}
\end{equation}
with $0\le\mu\le 1$ describing fermions with positive mass all the way from
zero to infinity. For small $\mu$, $\mu$ is proportional to the fermion mass.
When $\hat H_a = \epsilon (\log(T_w))$, the bare mass $\mu$ is exactly the
standard mass term in the domain wall formalism~\cite{Shamir,Herbert3}. 
The analytical expressions derived in this paper 
are valid for any generic choice of $\hat H_a$.
Our numerical results are obtained with the choice of $\hat H_a =\epsilon(H_w)$.

The massless Overlap-Dirac operator has exact zero eigenvalues, due to
topology, in topologically non-trivial gauge fields, and these zero
eigenvalues are always paired with eigenvalues exactly equal to
unity~\cite{Herbert1,EHN}. All exact zero eigenvalues are chiral, and
their partners at unity are also chiral but with opposite
chirality. The rest of the eigenvalues of the Overlap-Dirac operator
come in complex conjugate pairs and one eigenvector is obtained by the
operation of $\gamma_5$ on the other eigenvector. The behavior of the
Overlap-Dirac operator on the lattice is essentially that of a
continuum massless fermion.

The external fermion propagator is given by 
\begin{equation}
{\tilde D}^{-1}(\mu)=(1-\mu)^{-1}\left[D^{-1}(\mu) -1\right]~.
\end{equation}
The subtraction at $\mu=0$ is evident from the original overlap
formalism~\cite{over} and the massless propagator anti-commutes with
$\gamma_5$~\cite{Herbert1,Herbert2}. With our choice of subtraction
and overall normalization the propagator satisfies the relation
\begin{equation}
\mu \langle b^\dagger | \Bigl[ \gamma_5{\tilde D}^{-1}(\mu) \Bigr]^2
| b \rangle
= \langle b^\dagger | {\tilde D}^{-1}(\mu) | b \rangle
\ \ \ \ \forall \ \ b \ \ \ {\rm satisfying} \ \ \ 
\gamma_5 | b \rangle= \pm | b\rangle
\label{eq:Goldstone}
\end{equation}
for all values of $\mu$ in an arbitrary gauge field background.  This
relation will be evident from the discussion in
section~\ref{sec:chiral_cond}.  The fermion propagator on the lattice
is related to the continuum propagator by
\begin{equation}
D_c^{-1} (m_q) = Z^{-1}_\psi {\tilde D}^{-1}(\mu)
\qquad {\rm with} \quad m_q  = Z_m^{-1} \mu
\end{equation}
where $Z_m$ and $Z_\psi$ are the mass and wavefunction renormalizations,
respectively. Requiring that (\ref{eq:Goldstone}) hold in the continuum,
results in $Z_\psi Z_m =1$. For
$\hat H_a =\epsilon(H_w(m))$ we find
\begin{equation}
Z_\psi = Z_m^{-1} = 2m
\end{equation}
at tree level, and
a tadpole improved estimate gives
\begin{equation}
Z_\psi = Z_m^{-1} = {2 \over u_0} \left[ m - 4 (1 - u_0) \right] ~,
\end{equation}
where $u_0$ is one's favorite choice for the tadpole link value. Most
consistently, for the above relation, it is obtained from $m_c$, the
critical mass of usual Wilson fermion spectroscopy. 

In this paper, we study QCD in the quenched limit on the lattice using
the Overlap-Dirac operator. Our aim is to study the behavior of chiral
susceptibilities on a finite lattice in the massless limit.  In
section~\ref{sec:chiral_cond}, we derive analytic expressions for
various chiral susceptibilities using the spectral decomposition of
the Overlap-Dirac operator. These formulae do not reveal any new
physics not already known from formal arguments in the continuum;
however, they are valid for any lattice spacing, not only in the
continuum limit.  These formulae will help us simplify the numerical
computations and disentangle contributions due to global topology from
the rest of the contributions. We also discuss the massless limit of
the chiral susceptibilities in quenched QCD.  In
section~\ref{sec:details}, we present some details of the algorithm
used to numerically compute the chiral susceptibilities on a finite
lattice. Our numerical results are presented in
section~\ref{sec:results} and compared with the analytical expressions
obtained in section~\ref{sec:chiral_cond}. Similar numerical work has
been done recently using domain wall fermions with a finite extent in
the fifth direction~\cite{Col_lat98}.

\section{Chiral condensate and susceptibilities}
\label{sec:chiral_cond}

An analysis of the Overlap-Dirac operator can be performed in an
arbitrary gauge field background by noting that $\hat H_a$ can be
written in block diagonal form with $2\times 2$ blocks in the chiral
basis~\cite{EHN}.  We start by recalling a few properties of the
Overlap-Dirac operator $D$ for a single Dirac fermion: $(\gamma_5 D)^2$ 
commutes with $\gamma_5$ and both operators can therefore be
diagonalized simultaneously. The massive Overlap-Dirac operator can be
related to the massless one, from (\ref{eq:Dmu}), as
\begin{equation}
D(\mu)=\left(1-\mu \right) \left[ D(0) + {\mu\over 1-\mu}\right]\quad .
\label{eq:Dmu_0}
\end{equation}
We find it convenient to work in the chiral eigenbasis of $(\gamma_5 D(0))^2$
for massless fermions, $\mu = 0$. In this basis $D(\mu)$ takes on the following
block diagonal form~\cite{EHN}:
\begin{itemize}
\item There are $|Q|$ blocks of the form
$$\pmatrix{ \mu & 0 \cr 0 & 1 \cr}$$ associated with the global
topology of the configuration. These $|Q|$ blocks will be robust under
perturbations of the gauge field background.  The two modes per block
have opposite chiralities. The chiralities of the modes with
eigenvalues $\mu$ will all be the same and dictated by the sign of the
global topology $Q$.
\item There are $2NV-|Q|$ blocks of the form
$$
\pmatrix{ (1-\mu) \lambda_i^2 + \mu & (1-\mu)\lambda_i\sqrt{1-\lambda_i^2} \cr
-(1-\mu)\lambda_i\sqrt{1-\lambda_i^2} & (1-\mu)\lambda_i^2 + \mu \cr } $$
with $0 \le \lambda_i \le 1$, $i=1,2,\cdots,2NV-|Q|$, depending on the
background gauge field. The $\lambda_i^2$'s are the eigenvalues of
$(\gamma_5D(0))^2$.
We leave open the possibility that $n_z^\prime$
of these eigenvalues could be exactly zero. These are accidental zeros
in that they would be lifted by a small perturbation of the gauge field
background. In each subspace, $\gamma_5 = \pmatrix { 1 & 0 \cr 0 & -1 \cr}$.
\end{itemize}
The matrix can be inverted in the above form to give the following block
diagonal form for the external propagator 
${\tilde D}^{-1}(\mu)=(1-\mu)^{-1}\left[D^{-1}(\mu) -1\right]$:
\begin{itemize}
\item There are $|Q|$ blocks of the form
$$\pmatrix{ {1 \over \mu} & 0 \cr 0 & 0 \cr}$$
\item There are $2NV-|Q|$ blocks of the form
$${1\over \lambda_i^2(1-\mu^2) + \mu^2}
\pmatrix{ \mu(1-\lambda_i^2) 
& - \lambda_i\sqrt{1-\lambda_i^2} \cr
\lambda_i\sqrt{1-\lambda_i^2} &  \mu(1-\lambda_i^2)\cr } $$
\end{itemize}

We will use $\langle \cdots \rangle_A$ to denote the expectation value of
a fermionic observable in a fixed gauge field background, and $\langle \cdots
\rangle$ for the expectation value averaged over a gauge field ensemble. 
From the above expression for the fermion propagator in a background
gauge field, we obtain
\begin{equation}
{1\over V} \sum_{x} \langle \bar\psi(x)\gamma_5\psi(x) \rangle_A 
= {1\over V} {\rm Tr}[\gamma_5{\tilde D}^{-1}]= {Q\over \mu V}
\label{eq:parity}
\end{equation}
in a gauge field background with
global topology $Q$. Here ${\rm Tr}$ denotes a trace over space, color and
Dirac indices.
$\langle \bar\psi \gamma_5 \psi \rangle$ goes to zero when averaged over
all gauge field configurations.
We also find
\begin{equation}
{1\over V} \sum_{x} \langle \bar\psi(x) \psi(x) \rangle_A 
= {1\over V} {\rm Tr}[{\tilde D}^{-1}] = 
{|Q|\over \mu V} +
{1\over V} \sum_i {2\mu(1-\lambda_i^2)\over 
\lambda_i^2(1-\mu^2) + \mu^2} ~.
\label{eq:chiral}
\end{equation}
To discuss the gauge field average of $\langle \bar\psi \psi \rangle$
in the massless limit, we have to distinguish four cases
\footnote{ We do not consider the case of one flavor QCD since the chiral
symmetry is broken by the anomaly.}:
\begin{itemize}
\item We consider full QCD with $n_f > 1$ flavors in a fixed finite
volume. Gauge fields with non-zero topology are then suppressed by
$\mu^{n_f |Q|}$, the contribution from the zeromodes to $\det D(\mu)$.
Accidental zeromodes are also suppressed by the fermion determinant,
by $\mu^{2 n_f n_z^\prime}$.
Therefore, we find from eq.~(\ref{eq:chiral})
\begin{equation}
{1\over V} \sum_{x} \langle \bar\psi(x) \psi(x) \rangle = {\cal O}(\mu)
\end{equation}
and the chiral condensate vanishes in the massless limit in a finite volume.
\item We consider full QCD with $n_f > 1$ flavors in the infinite volume
limit, taking the volume to infinity before taking the mass to zero.
In this limit we expect $\langle |Q| \rangle$ to be proportional to
$\sqrt{V}$, and therefore the first term in (\ref{eq:chiral}) will go
to zero in the infinite volume limit.
The spectrum, $\lambda_i$, becomes continuous in the infinite
volume limit, represented by a spectral density, $\rho(\lambda)$. Then
we obtain,
\begin{equation}
\lim_{\mu\rightarrow 0}\lim_{V\rightarrow \infty}{1\over V} 
\sum_x \langle \bar\psi(x)\psi(x) \rangle = \pi\rho(0)
\label{eq:cond}
\end{equation}
and the chiral condensate is proportional to the spectral density at zero
just as expected in the continuum~\cite{Banks}.
\item We consider a quenched theory in a fixed finite volume.
Now, gauge field configurations with nontrivial topology and accidental
zeromodes are not suppressed by a fermion determinant, and we obtain
\begin{equation}
\lim_{\mu\rightarrow 0}{1\over V} \sum_{x} \langle \bar\psi(x) \psi(x) \rangle
= \lim_{\mu\rightarrow 0} {\langle |Q|+2n_z^\prime \rangle \over \mu V}
\label{eq:chiral_massless}
\end{equation}
$2 n_z^\prime$ is the number of accidental, paired zero modes. Since they
are lifted by a small perturbation of the gauge fields, the gauge fields
for which such accidental zero modes exist are probably of measure zero,
and we expect $\langle n_z^\prime \rangle = 0$. Thus
in any finite volume $\langle \bar\psi \psi \rangle$ diverges like
${1\over \mu}$ in the quenched theory with a coefficient equal to
${\langle |Q| \rangle \over  V}$. The expectation value here is over a
pure gauge ensemble, and the coefficient is therefore expected to be
proportional to $1/\sqrt{V}$. 
\item We consider a quenched theory in the infinite volume limit, taking
the volume to infinity before taking the mass to zero. Then, the first
term in (\ref{eq:chiral}) vanishes and from the second we obtain
\begin{equation}
\lim_{\mu\rightarrow 0}\lim_{V\rightarrow \infty}{1\over V} 
\sum_x \langle \bar\psi(x)\psi(x) \rangle = \pi\rho(0)
\label{eq:finite_rho0}
\end{equation}
where $\rho(0)$ is the spectral density at zero in a pure gauge ensemble.
\end{itemize}

A two flavor theory has a flavor SU(2)$_V \times$SU(2)$_A \cong $O(4)
symmetry. Consider the (real) O(4) vector made up of 
\begin{equation}
\vec\phi = (\pi^a=
i\sum_{ij} \bar\psi_i(x)\gamma_5\tau^a_{ij}\psi_j(x) ,
 ~ f_0=\sum_i\bar\psi_i(x)\psi_i(x))
\end{equation}
where $f_0$ is also known as $\sigma$,
and the opposite parity (real) O(4) vector made up of 
\begin{equation}
\vec{\tilde\phi} = (a_0^a= -\sum_{ij} \bar\psi_i(x)\tau^a_{ij}\psi_j(x) ,
 ~ \eta=i\sum_i\bar\psi_i(x)\gamma_5\psi_i(x)) ~.
\end{equation} 
Here $\tau^a$, $a=1,2,3$, are the SU(2) generators.  All eight
components are invariant under a global U(1) rotation associated with
fermion number.  Under a vector SU(2) flavor rotation, the $\pi^a$
rotate among themselves, the $a_0^a$ rotate among themselves, and
$f_0$ and $\eta$ are left invariant. Under an axial SU(2) flavor
rotation, the $\pi^a$ mix with $f_0$ and the $a_0^a$ mix with $\eta$.
Finally, under a global U(1) axial rotation, $\vec\phi$ mixes with
$\vec{\tilde\phi}$.

Consider the space averaged pion propagator
\begin{equation}
\chi_\pi = \sum_{x,y} {1\over V} \langle \pi^a(x)\pi^a(y) \rangle =
{2\over V} \langle {\rm Tr}(\gamma_5\tilde D)^{-2} \rangle ~.
\label{eq:pion1}
\end{equation}
$(\gamma_5\tilde D)^{-2}$ is diagonal in the chiral basis.
There are $|Q|$ eigenvalues equal to ${1\over\mu^2}$ and $|Q|$ zero
eigenvalues from the global topology. The non-zero eigenvalues appear
in multiples of two and are equal to $(1-\lambda_i^2)/ 
\left( \lambda_i^2(1-\mu^2) + \mu^2 \right)$ which is nothing but
the diagonal entries of ${1\over\mu}{\tilde D}^{-1}$.
Relation (\ref{eq:Goldstone}) in section~\ref{sec:intro} therefore follows.
We also obtain
\begin{eqnarray}
{1\over V} {\rm Tr}(\gamma_5\tilde D)^{-2}
&=& {|Q|\over\mu^2 V} + {1\over V} \sum_i
{2(1-\lambda_i^2) \over \lambda_i^2(1-\mu^2) + \mu^2}  \cr
&=& {1\over \mu} \langle \bar\psi \psi \rangle_A \quad .
\label{eq:susp2}
\end{eqnarray}
Therefore, we find the relation
\begin{equation}
\chi_\pi =  \sum_{x,y} {1\over V} \langle \pi^a(x)\pi^a(y) \rangle =
{2\over\mu} \langle \bar\psi \psi \rangle.
\label{eq:pion2}
\end{equation}
We note that this relation holds for any volume and $\mu$, configuration
by configuration. In fact, from (\ref{eq:Goldstone}) it follows that
the relation holds for any chiral random source that might be used for
stochastic estimates of $\chi_\pi$ and $\langle \bar\psi \psi \rangle$.

Using, similarly,
\begin{equation}
{1\over V} {\rm Tr}({\tilde D}^{-2})
= {|Q|\over \mu^2V} + {1\over V} \sum_i
{2(1-\lambda_i^2) \left[ \mu^2(1-\lambda_i^2) -\lambda_i^2 \right]
\over [\lambda_i^2(1-\mu^2) + \mu^2]^2}
\label{eq:susp1}
\end{equation}
we find for the space averaged $a_0$ propagator
\begin{equation}
\chi_{a_0} = \sum_{x,y} {1\over V} \langle a_0^a(x) a_0^a(y) \rangle =
- {2\over V} \langle {\rm Tr}\tilde D^{-2} \rangle =
2 \langle {d\over d\mu} \langle \bar\psi \psi \rangle_A \rangle ~.
\label{eq:a_0}
\end{equation}
For the space averaged $f_0$ propagator we find
\begin{equation}
\chi_{f_0} = \sum_{x,y} {1\over V} \langle f_0(x) f_0(y) \rangle =
{4\over V} \langle [{\rm Tr}{\tilde D}^{-1}]^2 \rangle -
{2\over V} \langle {\rm Tr}\tilde D^{-2} \rangle ~,
\label{eq:f_0}
\end{equation}
and for the space averaged $\eta$ propagator
\begin{equation}
\chi_\eta = \sum_{x,y} {1\over V} \langle \eta(x) \eta(y) \rangle =
{2\over V} \langle {\rm Tr}(\gamma_5\tilde D)^{-2} \rangle -
{4\over V} \langle [{\rm Tr}(\gamma_5\tilde D)^{-1}]^2 \rangle ~.
\label{eq:eta}
\end{equation}

To discuss these chiral susceptibilities, we again distinguish four cases:
\begin{itemize}
\item We consider full QCD in a finite volume and recall that gauge fields
with non-zero topology are suppressed by $\mu^{2 |Q|}$, while gauge fields
with accidental zeromodes are suppressed by $\mu^{4 n_z^\prime}$ from the
fermion determinant. Using eqs.~(\ref{eq:susp2}), (\ref{eq:susp1}) and
(\ref{eq:parity}) we find
\begin{equation}
\chi_\eta - \chi_{a_0} = {8 \over V} \Biggl\langle \sum_i
{\mu^2(1-\lambda_i^2)^2 
\over [\lambda_i^2(1-\mu^2) + \mu^2]^2} \Biggr\rangle
+{4 \langle |Q| \rangle\over \mu^2 V} 
-{4 \langle Q^2 \rangle\over \mu^2 V} ~.
\label{eq:eta_a0}
\end{equation}
Taking the $\mu \to 0$ limit, the first term on the right hand side vanishes,
while for the others only the sectors with $Q = \pm 1$ contribute, for
which the two remaining terms cancel. Therefore we obtain
\begin{equation}
\lim_{\mu\rightarrow 0} \left( \chi_\eta - \chi_{a_0} \right) = 0 ~,
\end{equation}
and, as expected in a finite volume, the O(4) symmetry is unbroken in
the massless limit. On the other hand, consider
\begin{eqnarray}
\omega & = & \chi_\pi - \chi_{a_0} = 
{2\over\mu} \langle \bar\psi \psi \rangle
- 2 \langle {d\over d\mu} \langle \bar\psi \psi \rangle_A \rangle\cr
& = & 
{8 \over V} \Biggl\langle \sum_i
{\mu^2(1-\lambda_i^2)^2 
\over [\lambda_i^2(1-\mu^2) + \mu^2]^2} \Biggr\rangle
+{4 \langle |Q| \rangle  \over \mu^2 V} \cr 
& = & \chi_\eta - \chi_{a_0}
+ {4 \langle Q^2 \rangle \over \mu^2 V} 
\quad .
\label{eq:omega}
\end{eqnarray}
We have used (\ref{eq:pion2}) and (\ref{eq:a_0}) in the first line,
(\ref{eq:susp2}) and (\ref{eq:susp1}) in the second line and
(\ref{eq:eta_a0}) in the last line.
Since $\chi_\eta - \chi_{a_0}$ vanishes in the massless limit, we
obtain
\begin{equation}
\lim_{\mu\rightarrow 0} \omega =
  \lim_{\mu\rightarrow 0} {4 \langle Q^2 \rangle \over \mu^2 V}~.
\label{eq:anomaly}
\end{equation}
The U(1)$_A$ symmetry, therefore, remains broken in a finite volume, due
to topology. Only the sectors with $Q = \pm 1$ contribute in the massless
limit, giving a finite result. In fact, the four fermi operator
making up $\chi_\pi - \chi_{a_0}$ written explicitly in terms of left
and right handed spinors, is nothing but the 
t'Hooft vertex.

\item We consider full QCD in the infinite volume limit, taking the volume
to infinity first. Then, one expects the O(4) symmetry to be broken down
to O(3)$\cong$SU(2), where the pions are the Goldstone bosons associated
with the spontaneous symmetry breaking. Eq.~(\ref{eq:pion2}) is consistent
with this expectation, implying that $m_\pi^2 \propto \mu$. The $\eta$,
on the other hand, is expected to remain massive due to the axial U(1)
anomaly. Therefore we find, using (\ref{eq:eta}), (\ref{eq:susp2})
and (\ref{eq:parity}),
\begin{eqnarray}
0 &=& \lim_{\mu\rightarrow 0}\lim_{V\rightarrow \infty} \mu \chi_\eta =
\lim_{\mu\rightarrow 0}\lim_{V\rightarrow \infty} {\mu \over V}
\sum_{x,y} \langle \eta(x) \eta(y) \rangle \cr
&=& \lim_{\mu\rightarrow 0}\lim_{V\rightarrow \infty} \left\{
2 \langle \bar\psi \psi \rangle - {4 \langle Q^2 \rangle
\over \mu V} \right\}~,
\label{eq:etamass}
\end{eqnarray}
which leads to the relation
\begin{equation}
\lim_{\mu\rightarrow 0}\lim_{V\rightarrow \infty} {2 \langle Q^2 \rangle
 \over \mu V} =
\lim_{\mu\rightarrow 0}\lim_{V\rightarrow \infty} \langle \bar\psi
\psi \rangle ~,
\end{equation}
a relation that has been derived before in the continuum~\cite{Smilga}.

\item We consider a quenched theory in a fixed finite volume, assuming
two flavors of valence fermions. Now, gauge fields with non-zero
topology are not suppressed, and we obtain from (\ref{eq:eta_a0}) for
$\chi_\eta - \chi_{a_0}$ in the massless limit:
\begin{equation}
\lim_{\mu\rightarrow 0} \left( \chi_\eta - \chi_{a_0} \right) =
\lim_{\mu\rightarrow 0} {4\over \mu^2 V} 
\langle |Q| + 2 n_z^\prime - Q^2 \rangle ~.
\end{equation}
Though we expect $\langle n_z^\prime \rangle = 0$ for the number of accidental
paired zero modes, we find not only that the O(4) symmetry remains broken
in the massless quenched theory in a finite volume, but that $\chi_\eta -
\chi_{a_0}$ diverges in the massless limit, unless we restrict the
quenched theory to the gauge field sector that has $|Q|=0,1$.
%With this restriction, one can still see the global U(1) anomaly, but it
%diverges in the massless limit.
With this restriction, one can still see the global U(1) anomaly; however,
$\omega$ diverges as $4 \langle |Q| \rangle / (\mu^2 V)$
in the massless limit.

\item We consider a quenched theory in the infinite volume limit, taking
the volume to infinity before taking the mass to zero. From (\ref{eq:pion1})
and (\ref{eq:pion2}) it is clear that
\begin{equation}
\chi_\pi - \chi_\eta = \langle {4 Q^2 \over \mu^2 V} \rangle~.
\end{equation}
If we restrict ourselves to the topologically trivial sector $\pi$ and $\eta$
will be degenerate. If we allow all topological sectors, the right hand
side of the above equation will diverge in the massless limit and the
$\eta$ particle will not be well defined ($\chi_\eta$ will become negative).
\end{itemize}

\section{Numerical details}
\label{sec:details}

The main quantity that needs to be computed numerically is the fermion
propagator
$\tilde D^{-1}$. Certain properties of the Overlap-Dirac operator enable us
to compute the propagator for several fermion masses at one time using the
multiple Krylov space solver~\cite{Jegerlehner} and also go directly to the
massless limit. Our numerical procedure to compute the fermion bilinear
in (\ref{eq:chiral}) and $\omega$ in (\ref{eq:omega}) proceeds as follows.

\begin{itemize}
\item We note that 
\begin{equation}
H_o^2(\mu) = D^\dagger(\mu) D(\mu) = D(\mu) D^\dagger(\mu) = 
\left(1-\mu^2 \right) \left[ H_o^2(0) + {\mu^2\over 1-\mu^2} \right]
\label{eq:H_o_mu}
\end{equation}
with
\begin{equation}
H_o^2(0) = {1\over 2} + {1\over 4} \left[\gamma_5\epsilon(H_w) +
 \epsilon(H_w)\gamma_5 \right]
\label{eq:H_o_0}
\end{equation}
\begin{itemize}
\item Eq.~(\ref{eq:H_o_mu}) implies that we can solve the set of equations
$H_o^2(\mu) \eta(\mu) = b$ for several masses, $\mu$, simultaneously
(for the same right hand $b$) using
the multiple Krylov space solver described in
Ref.~\cite{Jegerlehner}. We will refer to this as the outer conjugate
gradient inversion.
\item It is now easy to see that $[H_o^2(\mu),\gamma_5]=0$,
implying that one can work with the source $b$ and solutions
$\eta(\mu)$ restricted to one chiral sector. 
\end{itemize}
\item The numerically expensive part of the Overlap-Dirac operator is the
action of $H_o^2(0)$ on a vector since it involves the action of
$[\gamma_5\epsilon(H_w) + \epsilon(H_w)\gamma_5]$ on a vector. If the
vector $b$ is chiral ({\it i.e.} $\gamma_5 b = \pm b$) then, 
$[\gamma_5\epsilon(H_w) + \epsilon(H_w)\gamma_5] b  = [\gamma_5 \pm 1]
\epsilon(H_w)b$. Therefore we only need to compute the action of
$\epsilon(H_w)$ on a single vector.
\item In order to compute the action of $\epsilon(H_w)$ on a vector one needs
a representation of the operator $\epsilon(H_w)$. For this purpose we use the
optimal rational approximation for $\epsilon(H_w)$ described in Ref.~\cite{EHN}.
As described in \cite{EHN}, the action of $\epsilon(H_w)$ in the optimal
rational approximation amounts to solving equations of the form
$(H_w^2 + q_k) \eta_k = b$
for several $k$ which again can be done efficiently using the multiple Krylov
solver~\cite{Jegerlehner}. We will refer to this as the inner
conjugate gradient inversion.
\item A further improvement can be made by projecting out a few low lying eigenvectors
of $H_w$ before the action of $\epsilon(H_w)$. The low lying eigenvectors can
be computed using the Ritz functional~\cite{Ritz}. This effectively reduces the
condition number of $H_w$ and speeds up the solution of $(H_w^2 + q_k) \eta_k = b$.
In addition the low lying subspace of $H_w$ is treated exactly in the action of
$\epsilon(H_w)$ thereby improving the approximation used for  $\epsilon(H_w)$.
\item Since in practice it is the action of $H_o^2(\mu)$ on a vector
we need, we can check for the convergence of the complete operator at
each inner iteration of $\epsilon(H_w)$. This saves some small amount
of work at $\mu=0$ and more and more as $\mu$ increases, while at $\mu=1$
(corresponding to infinitely heavy fermions) no work at all is required.
\end{itemize}

We stochastically estimate ${\rm Tr}\tilde D^{-1}$,
${\rm Tr}(\gamma_5\tilde D)^{-2}$ and 
${\rm Tr}\tilde D^{-2}$ in a fixed gauge field background.
To be able to do this efficiently, and in order to compute this at arbitrarily
small fermion masses, we first compute the low lying spectrum of 
$\gamma_5 D(0)$ using the Ritz functional method~\cite{Ritz} . This gives us,
in particular, information about the number of zero modes and their 
chirality.\footnote{The
topology and chirality information can also be obtained from the spectral
flow method of $H_w$~\cite{specflow}.}
In gauge fields with zero modes we always find
that all $|Q|$ zero modes have the same chirality. We have not found
any accidental zero mode pairs with opposite chiralities.
We then stochastically estimate the three traces above in the chiral sector
that does not have any zero modes. In this sector, the propagator with any
source is non-singular even for zero fermion mass. Given a Gaussian random
source $b$ with a definite chirality we solve the equation
$H_o(\mu)^2 \eta(\mu) = b$ for several values of $\mu$. Then we compute
$(1-\mu)\xi(\mu) = D^\dagger(\mu) \eta(\mu) - b$. The stochastic estimate of
${\rm Tr}\tilde D^{-1}$ is obtained by computing $b^\dagger \xi(\mu)$ and
averaging over several independent Gaussian chiral vectors $b$.
The stochastic estimate of ${\rm Tr}(\gamma_5\tilde D)^{-2}$ 
is similarly obtained by computing $\xi^\dagger(\mu)\xi(\mu)$ and 
averaging over several independent Gaussian chiral vectors $b$.
To estimate ${\rm Tr}\tilde D^{-2}$ we also compute 
$(1-\mu)\xi^\prime(\mu) = D(\mu) \eta(\mu) - b$. Then ${\rm Tr}\tilde D^{-2}$ is
stochastically obtained by computing ${\xi^\prime}^\dagger(\mu) \xi(\mu)$
and averaging over several independent Gaussian chiral vectors $b$.
Since the stochastic estimates were done in the chiral sector that does not
contain any zero modes, the result has to be doubled to account for the
contributions from the non-zero modes in both
chiral sectors. The contributions from the zero modes due to topology,
finally, are added on analytically from the appropriate
equations in section~\ref{sec:chiral_cond}. 
Since modes with $\lambda_i^2=1$ do not contribute at
all to $\tilde D^{-1}$, we do not have to worry about double counting these
edge modes.

\section {Numerical results}
\label{sec:results}

For our numerical simulations, we used the usual hermitian
Wilson--Dirac operator $H(m)= \gamma_5 W_w(-m)$ with a negative
fermion mass term~\cite{Herbert1} at $m=1.65$. Our main results are
for quenched $SU(3)$ $\beta=5.85$, where we have three volumes, and
$\beta=5.7$, where we considered one. In Table~\ref{tab:params}, we
show the simulation parameters and number of configurations used.
From our previous work with the spectral flow of the hermitian
Wilson--Dirac operator~\cite{specflow}, we have higher statistics
estimates of $\langle |Q| \rangle$ and $\langle Q^2\rangle$ which we
quote in Table~\ref{tab:params} with $N_{\rm top}$ configurations
used. For our stochastic estimates of the various traces, we used
$N_{\rm conf}$ configurations and $5$ stochastic Gaussian random
sources. We found $5$ sources to be sufficient. For the plots of
$\langle\bar\psi \psi\rangle/\mu$ and $\omega$, we used $N_{\rm conf}$
configurations for the topology term with all statistical errors
computed using a jackknife procedure.

\begin{figure}
\epsfxsize = \textwidth
%\epsfxsize=7in
%\centerline{\epsfbox[0 0 576 576]{pbp_notop.ps}}
\centerline{\epsffile{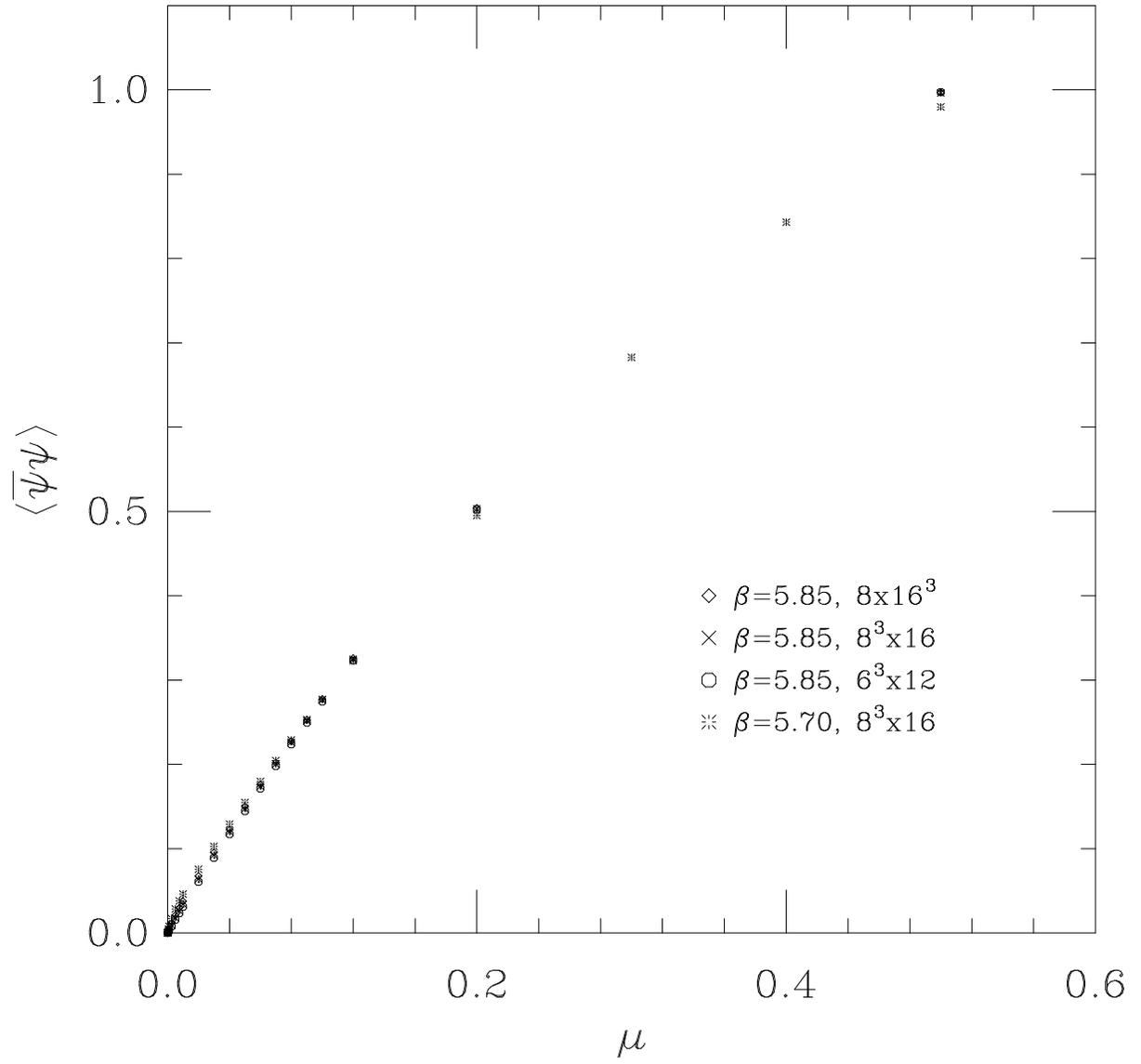}}
\caption{Plot of $\langle \bar\psi \psi \rangle$ without topology for the data
sets in Table~\ref{tab:params}.}
\label{fig:pbp_notop}
\end{figure}

\begin{figure}
\epsfxsize = \textwidth
%\epsfxsize=5in
%\centerline{\epsfbox[50 40 600 600]{pbp_notop.ps}}
\centerline{\epsffile{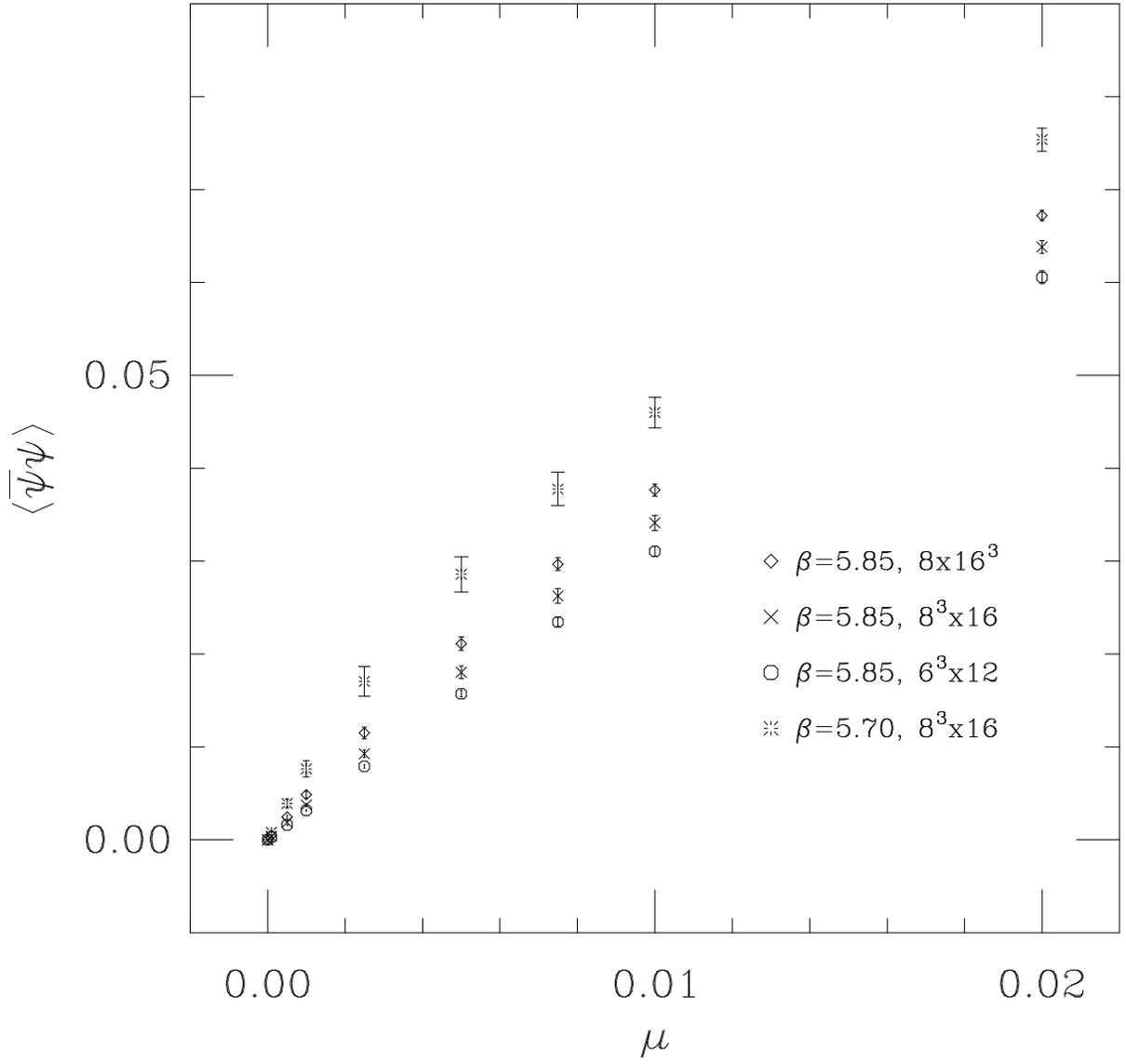}}
\caption{Expanded view of $\langle \bar\psi \psi \rangle$ without
topology.}
\label{fig:pbp_notop_expanded}
\end{figure}

\begin{figure}
\epsfxsize = \textwidth
\centerline{\epsffile{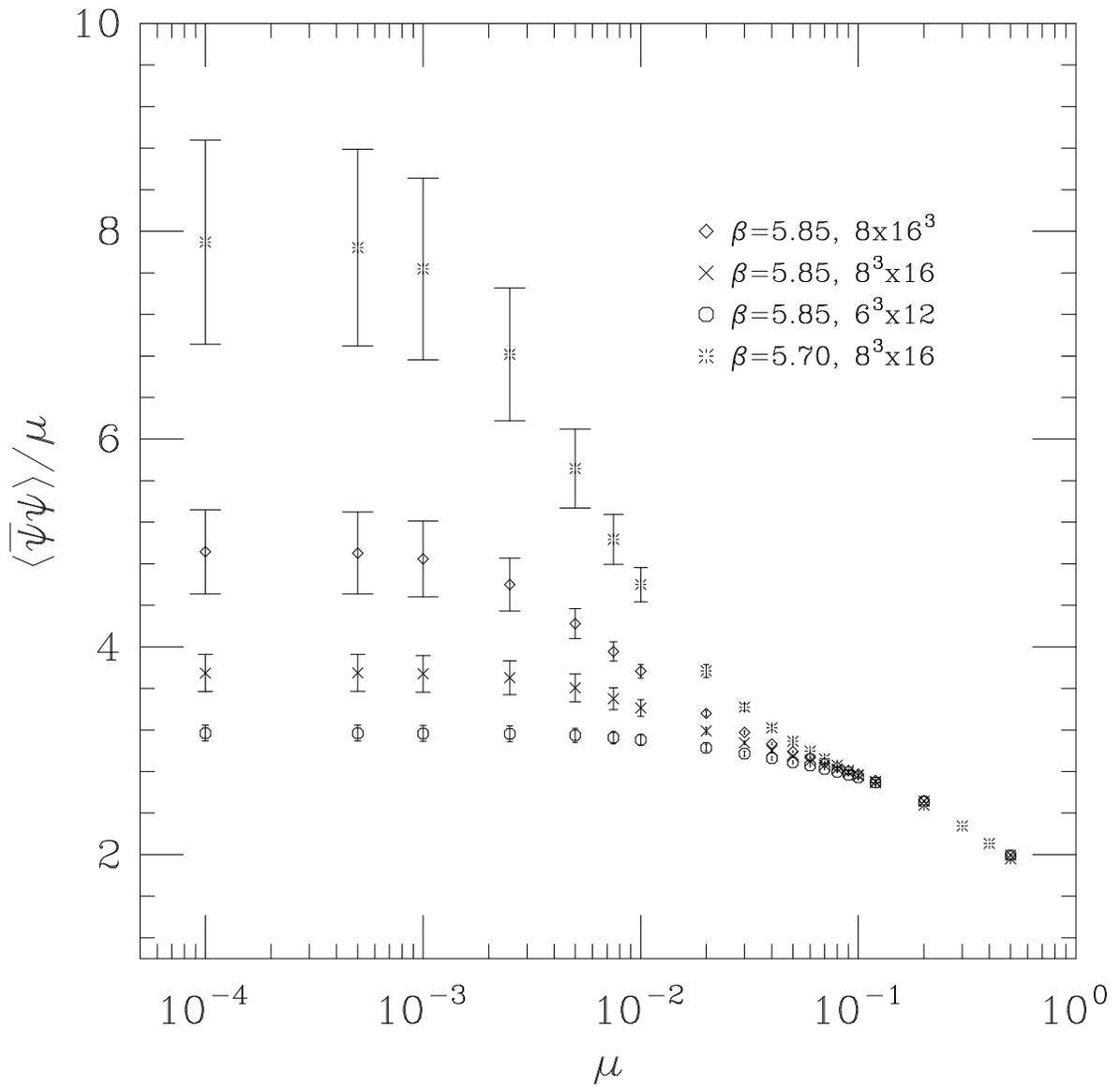}}
\caption{Plot of $\langle \bar\psi \psi \rangle/\mu$ without topology. 
Strong finite volume effects are manifested as constancy in
$\mu$. The mass region where finite volume effects sets in moves to
smaller mass values as the volume is increased.
}
\label{fig:pbp_mu}
\end{figure}

The fermion masses we used for our simulations range from $0$ to
$0.5$. We note that the number of (outer) conjugate gradient
iterations we require is typically about $50$, $90$, and $200$ with a
convergence accuracy of $10^{-6}$ (normalized by the source) for our
three volumes (in increasing order) at $\beta=5.85$. This is
consistent with our expectation that while we have chosen a chiral
source that has no overlap with the zero modes, the smallest non-zero
eigenvalues of $H_o^2(0)$ are decreasing with increasing volume. The
number of inner conjugate gradient iterations (used for the
application of $\epsilon(H_w)$ on a vector) is about 100, 105, and 150
with a convergence accuracy of $10^{-6}$ for our three volumes at
$\beta=5.85$. We projected out the 10, 10, and 20 lowest eigenvectors
of $H_w$ in the computation of $\epsilon(H_w)$, respectively, before
using the optimal rational approximation to compute $\epsilon(H_w)$ in
the orthogonal subspace.  While costly, the pole
method~\cite{EHN,Herbert4} of implementing $\epsilon(H_w)$ is clearly
the most cost effective in terms of floating point operations compared
to other implementations of the Overlap--Dirac
operator~\cite{EHN,Borici},

We show our result for $\langle\bar\psi \psi\rangle$ without the topology
term included in Figure~\ref{fig:pbp_notop}. Shown (but not discernible) are
the results for all four volumes. An expanded view for small mass is shown in
Figure~\ref{fig:pbp_notop_expanded}. For small $\mu$, a linear dependence on
$\mu$ with zero intercept is expected from Eq.~(\ref{eq:chiral}).
If there is to be a chiral condensate in the infinite volume limit the
slope near $\mu=0$ should increase with volume. To see this effect
we show $\langle\bar\psi \psi\rangle/\mu$ for all our data sets in
Figure~\ref{fig:pbp_mu}. 
For sufficiently small $\mu$, we see linearity in $\mu$ due
to finite volume effects. This linearity is manifested as a constancy in 
$\langle\bar\psi \psi\rangle/\mu$. As the volume is increased, the
mass region where linearity sets in is shifted to smaller values.
However, the slope near $\mu=0$ does not increase proportional to the
volume and we conclude that there is no definite evidence for a
chiral condensate in the infinite volume limit yet.

\begin{figure}
\epsfxsize = \textwidth
\centerline{\epsffile{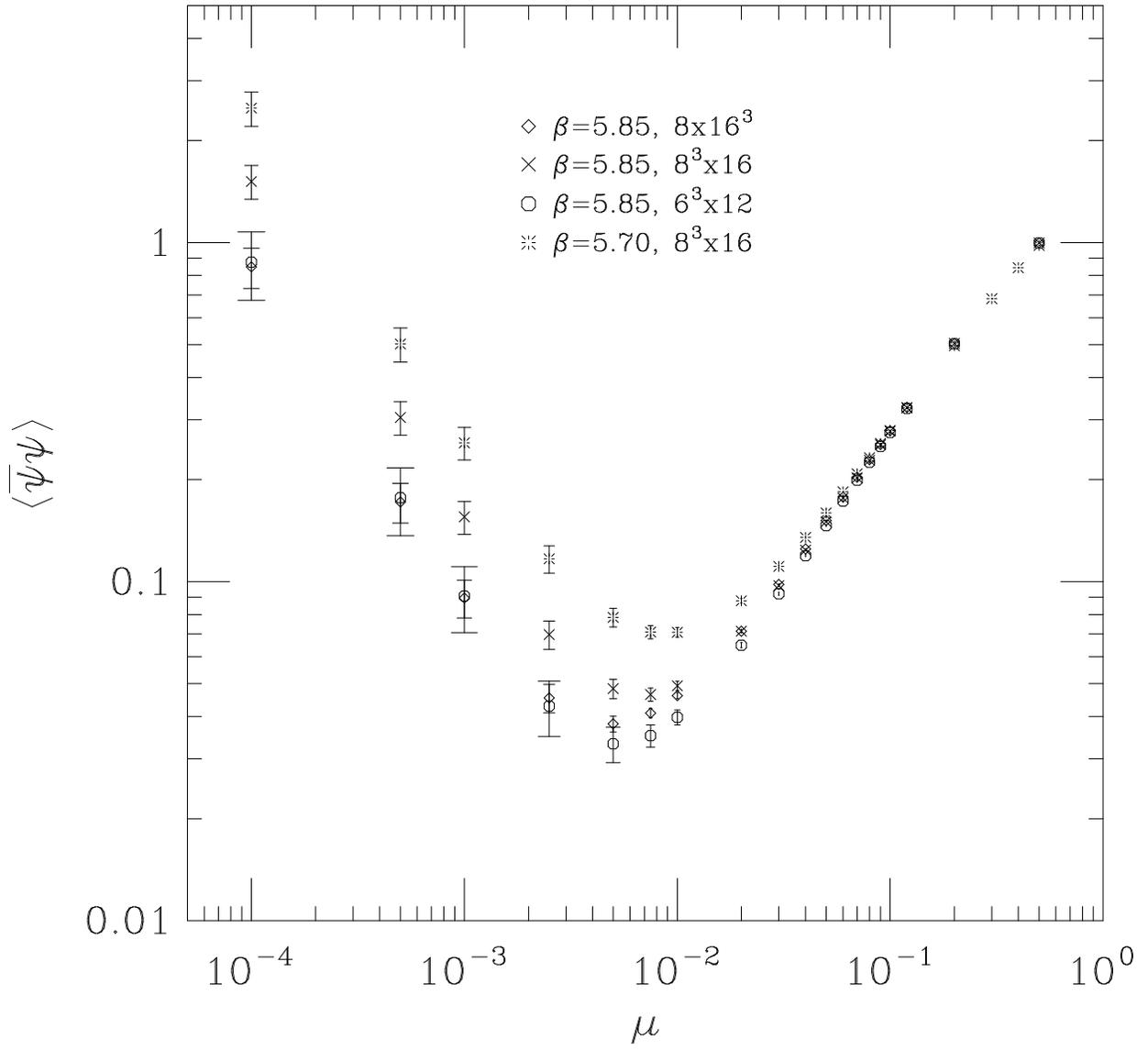}}
\caption{Plot of $\langle \bar\psi \psi \rangle$ with topology.}
\label{fig:pbp_withtop}
\end{figure}

We show $\langle\bar\psi \psi\rangle$ with the topology term included in
Figure~\ref{fig:pbp_withtop}. As expected from Eq.~(\ref{eq:chiral}),
the topology term diverges in a finite volume for $\mu\rightarrow
0$. However, the singularity is decreasing for increasing volume. For
sufficiently large volumes, we expect that the quenched $\langle|Q|\rangle$
scales like $\sqrt V$. From Table~\ref{tab:params}
we see that the increase of $\langle|Q|\rangle$ from
$8^3\times 16$ to $8\times 16^3$ is consistent with a $\sqrt V$ growth.
But this is not the case when we compare the value at
$6^3\times 12$ with the one at $8^3\times 16$. This is attributed
to a finite volume effect present at $6^3\times 12$~\cite{specflow}.

\begin{figure}
\epsfxsize = \textwidth
\centerline{\epsffile{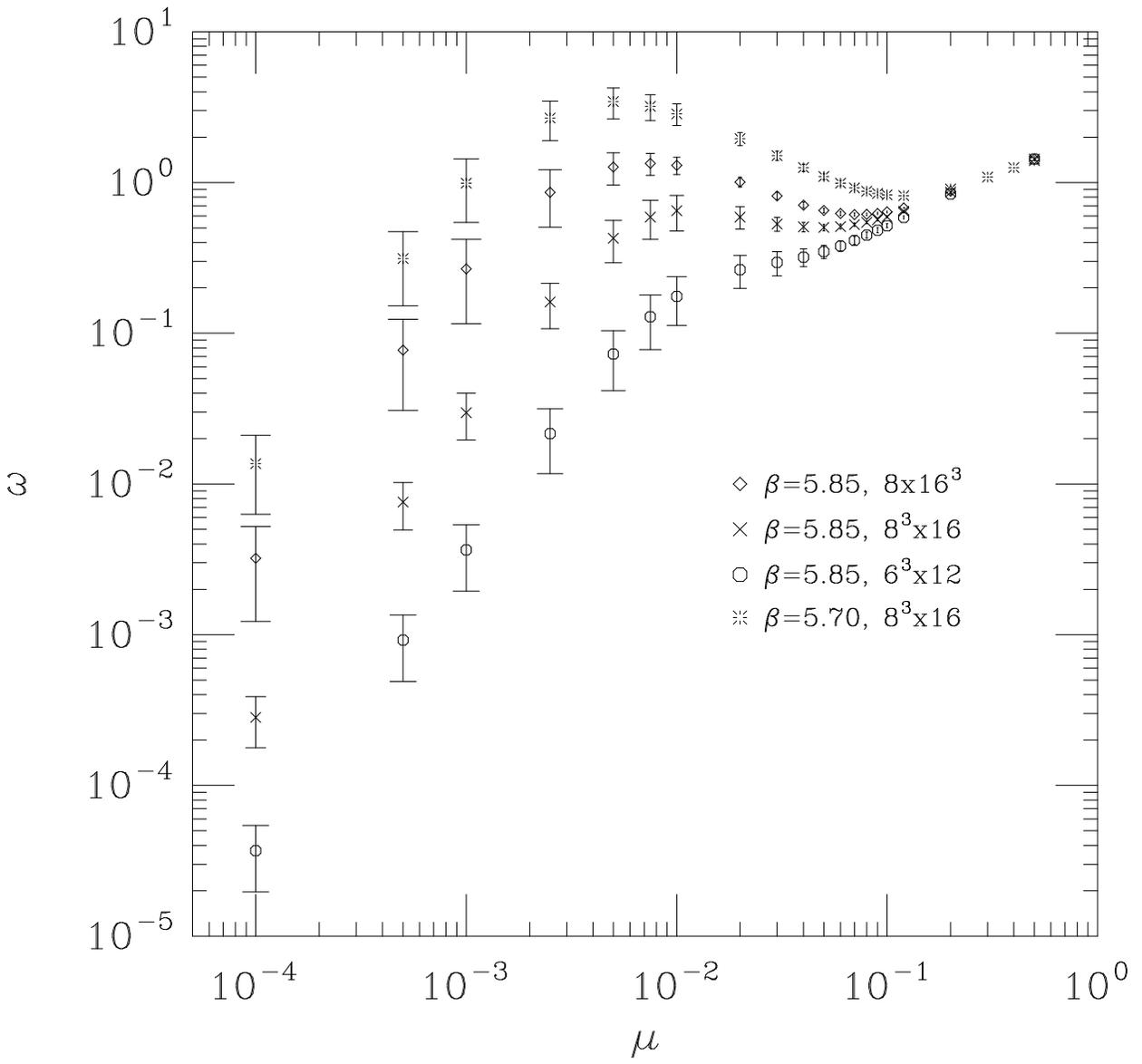}}
\caption{Plot of $\omega$ without topology.}
\label{fig:omega_notop}
\end{figure}

More insight into the possible onset of spontaneous chiral symmetry
breaking can be obtained by studying $\omega$ in
Eq.~(\ref{eq:omega}). If chiral symmetry is broken then $\omega$
should diverge as ${1\over\mu}$; otherwise, it should go to zero as
$\mu^2$. Evidence for chiral symmetry breaking in the infinite volume
limit should show up as a ${1\over\mu}$ behavior in $\omega$ at some
moderately small fermion masses ( an increase in the value of $\omega$
as one decreases the fermion mass) that turns over to a $\mu^2$
behavior at very small fermion masses.  We show $\omega$ without the
topology term added in Figure~\ref{fig:omega_notop}.
We see a turnover developing in the $\beta=5.85$ data as the
volume is increased around $\mu \sim 10^{-2}$.  The turnover is also
present in the data at $\beta=5.7$.  Therefore it is possible that
there is an onset of chiral symmetry breaking, but the effect is not
strong enough for us to be able to extract a value for the chiral
condensate.

\begin{figure}
\epsfxsize = \textwidth
\centerline{\epsffile{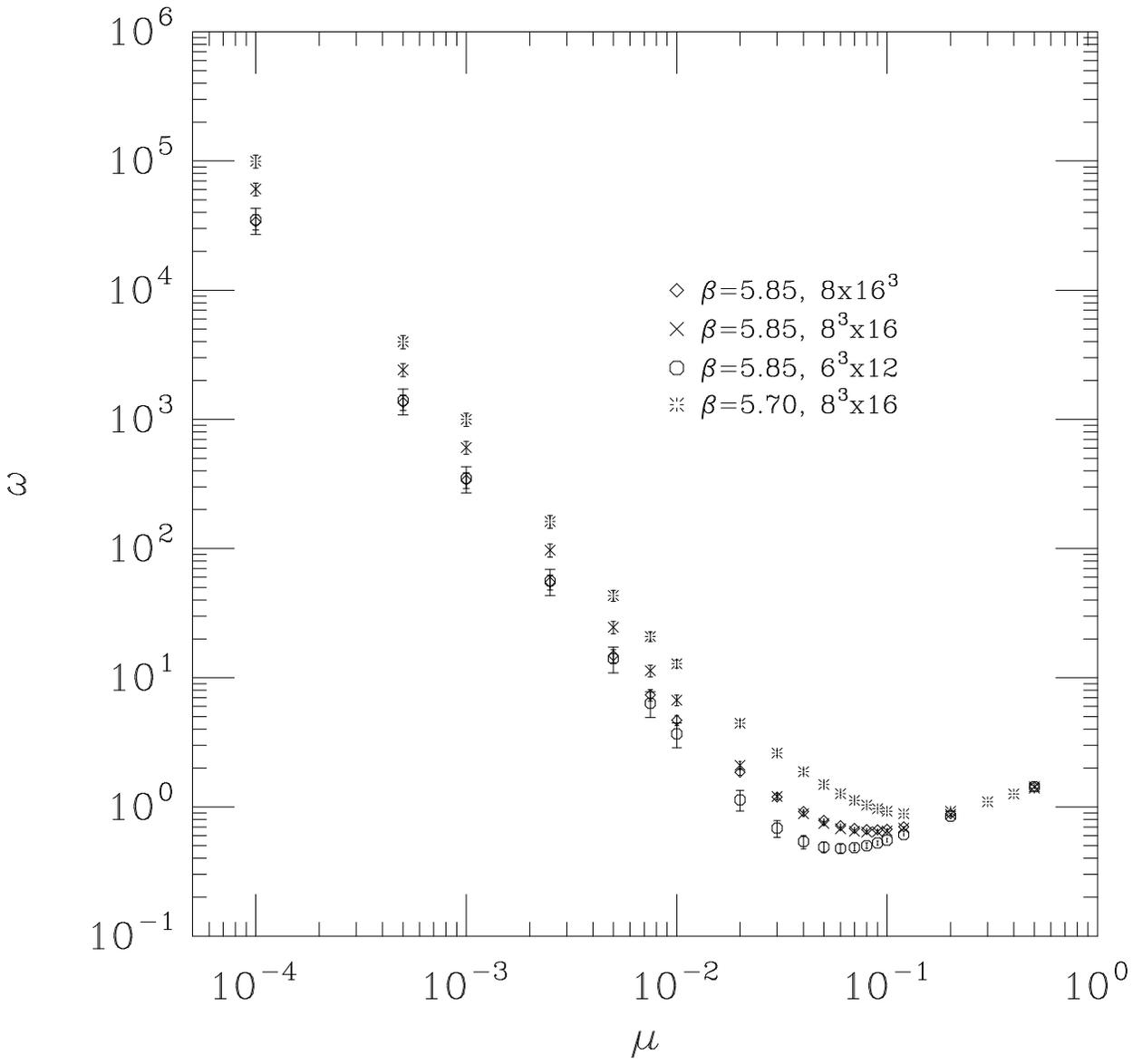}}
\caption{Plot of $\omega$ with topology.}
\label{fig:omega_withtop}
\end{figure}

When we add the topology term in $\omega$, as shown in
Figure~\ref{fig:omega_withtop}, the turnover region is completely
obscured. This result demonstrates how finite volume effects relevant
for a study of chiral symmetry are
obscured by topology. To be able to extract the chiral condensate
in a quenched theory from finite volume studies it would therefore
be helpful to remove the contribution from topology. This is possible
with the Overlap-Dirac operator as demonstrated here.

\section {Conclusions}
\label{sec:conclusions}

We have derived the standard continuum relations among various
fermionic observables by working on a finite volume lattice with the
Overlap-Dirac operator. Our results are general and apply equally well
to the Overlap--Dirac operator of Neuberger~\cite{Herbert1} and
domain wall fermions for infinite extent in the extra fifth
dimension~\cite{Herbert3}. These relations do not reveal any new
physics but allow us to disentangle the contribution of topology and
simplify our numerical simulations. We should emphasize that the
relations follow only because Overlap fermions satisfy the usual
chiral symmetries at any lattice spacing, not only in the continuum
limit.

We have shown that the standard relation between the pion
susceptibility and the fermion bilinear (c.f.~(\ref{eq:pion2})) is
properly reproduced by our implementation of the Overlap-Dirac
operator even in the chiral limit.  We found that for the volumes
studied the chiral condensate in the quenched approximation at small
fermion mass is dominated by the contribution from the global
topology. While this contribution eventually vanishes in the infinite
volume limit, it overpowers any signs for an onset of a non-vanishing
condensate for the volumes that we were able to study. Removing this
topological contribution, which is possible in our implementation of
the Overlap-Dirac operator, enabled us to perform a finite volume
analysis of chiral symmetry breaking. Still, our numerical simulations
for quenched $SU(3)$ gauge theory do not present strong evidence for a
chiral condensate in the infinite volume limit of the quenched theory;
however, there is some evidence for the onset of chiral symmetry breaking
as the volume is increased.

\acknowledgements 
This research was supported by DOE contracts DE-FG05-85ER250000 and
DE-FG05-96ER40979. Computations were performed on the workstation
cluster, the CM-2 and the QCDSP at SCRI, and the Xolas computing
cluster at MIT's Laboratory for Computing Science.
We would like to thank H.~Neuberger for discussions.

\newpage

\begin{table}
\begin{center}
\small
\begin{tabular}{|l|c|c|r|c|c|}
$\beta$ & size & $N_{\rm conf}$ & $N_{\rm top}$ & $\langle|Q|\rangle$
& $\langle Q^2\rangle$\\
\hline
$5.85$ & $8\times 16^3$ & 31 & 40 & $2.54\pm 0.35$ & $10.7\pm 3.5$ \\
$5.85$ & $8^3\times 16$ & 50 & 200 & $1.115\pm 0.067$ & $2.13\pm 0.22$ \\
$5.85$ & $6^3\times 12$ & 97 & 400 & $0.268\pm 0.025$ & $0.323\pm 0.037$ \\
\hline
$5.70$ & $8^3\times 16$ & 50 & 50 & $1.98 \pm 0.23$ & $6.6\pm 1.3$ \\
\end{tabular}
\end{center}
\caption{Simulation parameters at $m=1.65$. The number of
configurations $N_{\rm conf}$ was used for the trace estimates, and
$N_{\rm top}$ configurations were used for the estimates of
$\langle|Q|\rangle$ and $\langle Q^2\rangle$. The spectral flow method
of ref.~\protect\cite{specflow} was used for the estimates of
$\langle|Q|\rangle$ and $\langle Q^2\rangle$ for all but the first
line.}
\label{tab:params}
\end{table}

\end{document}